\documentclass[aps,preprint,floatfix,altaffilletter]{revtex4}

\usepackage[dvips]{graphicx}
\usepackage{color}
\usepackage{amsmath}
\usepackage{here}
\begin{document}
\begin{sloppy}

\title{Nanoscale plasmonic circulator}
\author{Arthur R. Davoyan}
\email{davoyan@seas.upenn.edu}
\address{Department of Electrical and Systems Engineering, University of Pennsylvania, Philadelphia,
Pennsylvania 19104, USA }
\author{Nader Engheta}
\email{engheta@ee.upenn.edu}
\address{Department of Electrical and Systems Engineering, University of Pennsylvania, Philadelphia,
Pennsylvania 19104, USA }

%  \date{\today}

%\email{davoyan@seas.upenn.edu}

\begin{abstract}
Here, we propose a conceptual approach for design of an ultracompact nanoscale passive optical circulator based on the excitation of plasmonic resonances. We study a three-port Y-junction with a deep subwavelength plasmonic nanorod structure integrated into its core. We show theoretically that such a structure immersed in the magneto-optical media may function as magnetically tunable scatterer tilting and rotating its near-field distribution and corresponding radiation. We demonstrate, using numerical simulations, that such a rotation of the near-field radiation yields a break in the symmetry of the coupling between the junction arms and the structure in such a way that the signal launched from any of the three ports is mostly transmitted into the next port in the circular order, while the other port is essentially isolated, thus providing the functionality of an optical circulator with subwavelength dimensions.

Keywords: Plasmonic nanodevices, optical isolation, nonreciprocal power flow circulation, near-field manipulation
\end{abstract}

%PACS number(s): {42.70.Qs, 78.20.Bh, 42.15.-i, 42.25.Bs, 45.20.Jj}
%\pacs{42.70.Qs, 78.20.Bh, 42.15.-i, 42.25.Bs, 45.20.Jj}
\maketitle

Advances in nanophotonics and plasmonic optics in recent years have provided miniaturization of optical devices and components suitable for on-chip integration~\cite{Int_opt_book,Joannopoulos,Bozh_rev}. In particular, some of the basic passive and active nanoscale optical integrated circuit elements including interconnects, beam splitters, switches and other components have been demonstrated recently~\cite{Bozh}. At the same time, fully functional signal processing in nanophotonics requires design of {\em nonreciprocal} elements, such as isolators or circulators (analogous, for example, with advances in microwave technology~\cite{Helszajn_1,Helszajn_2}).

Such nonreciprocal characteristics are achieved with materials exhibiting time-reversal symmetry breaking, such as magneto-optic materials (MO). (Other possibilities include utilizing nonlinearities~\cite{Shadrivov_diode,Miroshnichenko_diode,Assanto_diode} or materials having time dependent properties~\cite{Fan_isolator,Fan_isolator_v2}). However, at optical frequencies the magneto-optical response is usually weak, which results in a bulky and inefficient design of basic nonreciprocal optical components, not usually compatible with on-chip integration. For example, ``conventional" three port optical circulators - devices that guide light from any of the input ports into the next port in a circular order~\cite{Fay} - are typically more than several hundred microns large (i.e. $\gg \lambda\simeq 1 \mu$m)~\cite{Takei,Sugimoto}.

Progress in silicon photonics and magneto-photonic crystals promises a substantial miniaturization of key nonreciprocal optical devices. For instance, an on-chip silicon-based optical isolation~\cite{Ross_isolator}, one-way edge state propagation~\cite{Soljacic_oneway,Chui_crystal}, and compact photonic crystal circulators comparable with the operating wavelength (i.e. $\geq \lambda$)~\cite{Fan_circ_v1,Fan_circ_v2,Smigaj,Soljacic_circ} have been suggested and demonstrated recently. However, further size reduction of conventional dielectric devices down to the subwavelength scale ($\leq\lambda$) encounters fundamental challenges due to the light diffraction limit.

This is where highly confined surface plasmons - electromagnetic fields at metal surfaces - can be utilized in order to extend the technology beyond the diffraction limit~\cite{Bozh_rev,Maier_rev}. Furthermore, strong light localization in plasmonic structures can lead to a considerable increase of the nonreciprocal response. The latter was successfully implemented resulting in a significant enhancement of Kerr and Faraday effects~\cite{Beloletov_TKMOE,Wang_coreshell,Jain_coreshell,Tomita_polarKerr,Garcia_nanodisk}.

In our other work~\cite{our_prl} we have shown that plasmonic nanorod structures possessing a rotational symmetry, when immersed in magneto-optical media, exhibit a strong azimuthal symmetry breaking. In particular, it has been suggested that rotational symmetry of the structure gives rise to an eigenmode degeneracy at some of the plasmonic resonances. (It is worth mentioning here that the study of axially symmetric plasmonic nanoclusters~\cite{nanoclusters_v1,nanoclusters_v2} and oligomers~\cite{oligomers_v1} with high quality resonances has emerged recently as a rapidly developing and exciting field.) These eigenmodes may be represented as counter rotating eigenstates forming an azimuthally standing wave pattern in the unbiased regime. In presence of magneto-optical activity a strong mode interaction occurs yielding a frequency splitting of these counter-rotating states~\cite{Fan_circ_v2} which can be described by a simple empirical formula: $\omega_{\pm} = \omega_0(1\pm C\alpha)$; where ${\pm}$ correspond to two counter-rotating states, $\omega_0$ is the resonant frequency at which the mode degeneracy is observed in the unbiased case, $\alpha$ is the strength of magneto-optical activity, and $C$ is the intermodal coupling constant depending on the mode configuration. Note that an analogous symmetry breaking mechanism between the counter-rotating states was employed to design a compact photonic crystal circulator in~\cite{Fan_circ_v1,Fan_circ_v2}. Our recent study of the light scattering from such nonreciprocal nanostrctures revealed that the symmetry breaking between the counter rotating states leads to a strong near-field power flow  circulation resulting in the tilt and rotation of the scattered radiation pattern~\cite{our_prl}.

\begin{figure}[h!]
  \begin{center}
      \includegraphics[width=0.7\columnwidth]{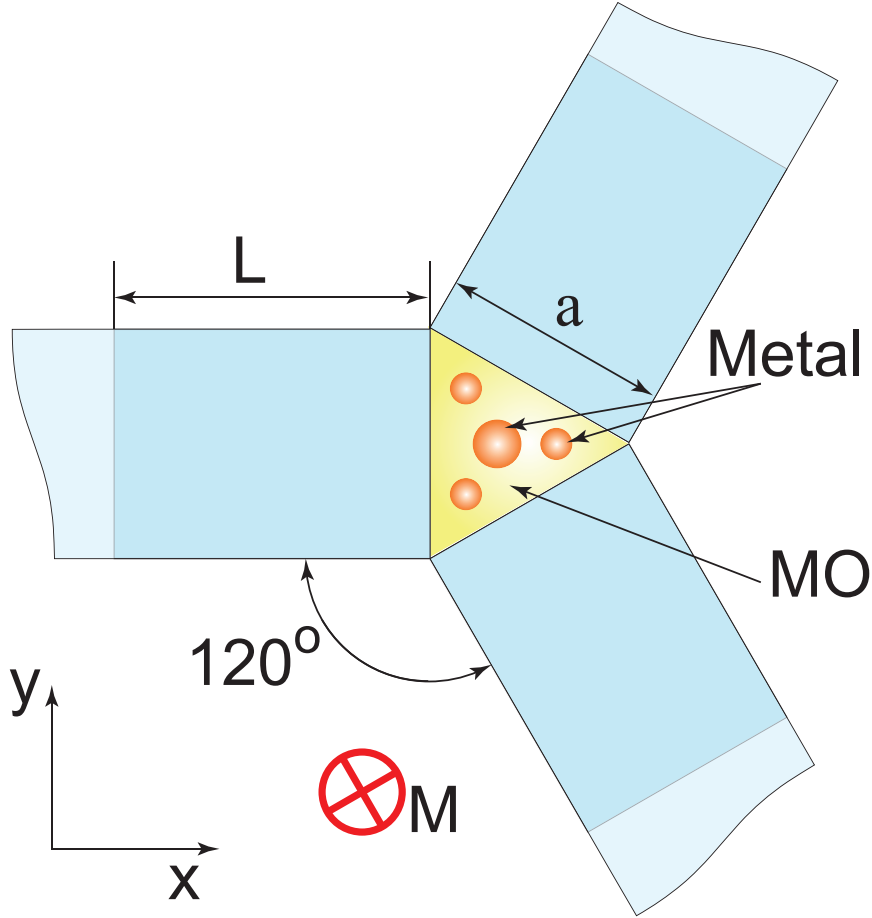}
      \caption{Schematic of our subwavelength three-port junction circulator. The detailed description of the structure and the parameters is given in the text.} \label{geometry}
  \end{center}
\end{figure}

In this Letter, based on our results from~\cite{our_prl} we propose a design for an ultracompact nanoscale plasmonic circulator, exploiting the above-mentioned mechanism of rotation of scattering from subwavelength plasmonic nanorod structure in the presence of MO materials. We study a three-port Y-junction with a plasmonic structure embedded in its core. We show that the structure acts as a resonant scatterer which enhances the junction's transmission characteristics. We reveal that introducing MO activity into the system yields rotation of the near-field scattering by the subwavelength structure, and numerically demonstrate that such rotation breaks the symmetry between the power transmitted through the junction's output arms. We show that with a relatively reasonable values of MO activity the signal launched into any of the ports is mostly guided into the next port (in the circular order), while the other output port is essentially isolated. Finally, we analyze influence of losses on the system operation and show that circulator performance remains high even in lossy regime.

Figure~\ref{geometry} shows the geometry of our proposed three-port junction circulator. For the sake of simplicity in the analysis and numerical simulations, and without loss of generality, we assume our system to be a two-dimensional (2D) structure. The circulator consists of three single-mode dielectric waveguides of width $a$ coupled with a plasmonic structure consisting of metallic nanorods with deeply subwavelength diameters immersed in MO medium as shown in Fig.~\ref{geometry}. We assume that the MO media is magnetized along the z-direction, and therefore, its relative permittivity function is described by an antisymmetric tensor:

 \begin{equation}
  \bar{\bar{\varepsilon}}= \left( \begin{array}{ccc}
   \varepsilon_{mo} & i\alpha & 0 \\
   -i\alpha & \varepsilon_{mo} & 0 \\
   0 & 0 & \varepsilon_\bot \end{array} \right), \nonumber
\end{equation}
where $\varepsilon_{mo}$ and $\varepsilon_\bot$ are diagonal components, $\alpha$ is the off-diagonal component responsible for the ``strength'' of MO (nonreciprocal) activity of the medium. The parameter $\alpha$ is usually very small at optical frequencies (of the order of $10^{-2}$ or less). For example, bismuth iron garnet (BIG) exhibits $\varepsilon_{mo}=6.25$, $\alpha=0.06$ and small optical losses~\cite{BIG}. In order to better understand the system dynamics of our proposed structure, in our theoretical studies we assume $\alpha$ ranging from zero to 0.15.

The plasmonic structure consists of a metallic nanorod with radius $R=50$nm surrounded by three other metallic nanorods with radii $R_s=10$nm as shown schematically in Fig.~\ref{geometry}. The distance between the centers of thinner and thicker nanorods is $80$nm. Here we assume the metal relative permittivity to be fixed at $\varepsilon_m=-10$, and, if not stated otherwise, the system is assumed to be lossless. Note that, for the sake of simplicity in our analysis and optimization we keep the material permittivities independent of frequency.  This does not introduce any limitation in our concept, since the dispersion of the material parameters can be easily included in the original design of our system.

This 4-nanorod  structure is placed in a uniform magnetized BIG host. At $\lambda \simeq 1.4 \mu$m and the parameters chosen above, this system exhibits a sharp plasmonic resonance, a strong near-field power flow circulation, and rotation of the near-field scattering at this resonance (not shown here, but in full accordance with our results shown in Ref.~\cite{our_prl}). Integrating this structure into the core of the three-port junction, as shown in Fig.~\ref{geometry}, we expect that the near-field circulation cause the asymmetric and nonreciprocal power redistribution into the output arms of the junction. In order to demonstrate our predictions we numerically study the circulator operation with a finite-element-based method in the COMSOL Myltiphysics$^{TM}$.

\begin{figure}[h!]
  \begin{center}
      \includegraphics[width=1\columnwidth]{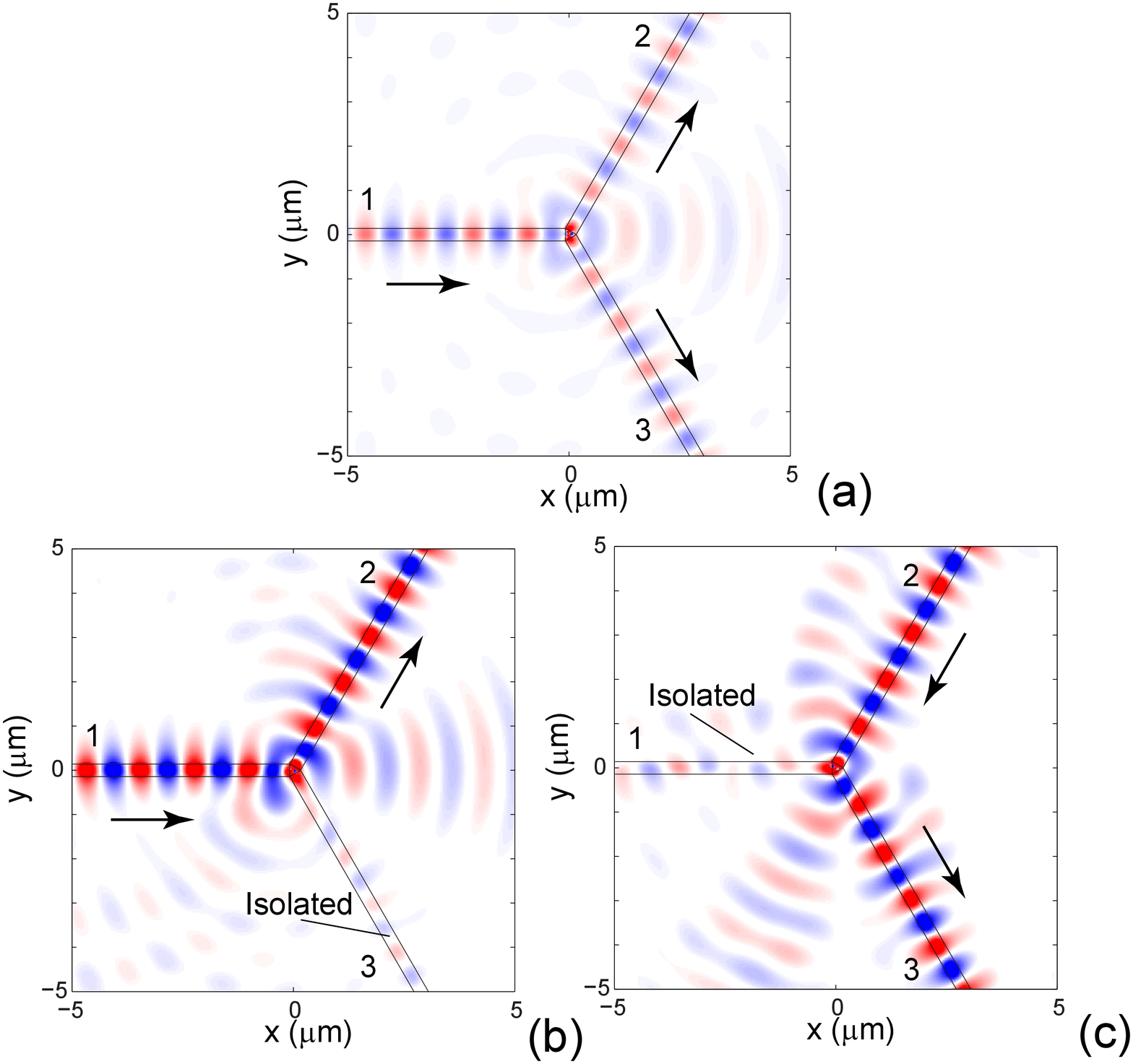}
      \caption{Simulation results for the distribution of the z-oriented magnetic field ($H_z$) in the circulator at the free-space wavelength $\lambda = 1.43\mu$m for a) unbiased ($\alpha = 0$), and b),c) biased ($\alpha=0.1$) systems, respectively. Arrows show the direction of the power flow. Numbers $1-3$ label the ports of the circulator.} \label{circulat}
  \end{center}
\end{figure}

We begin our analysis with a magnetically inactive system, i.e. unmagnetized case when $\alpha=0$. In this case no asymmetry or nonreciprocity is present and the system behaves as a symmetrical beam splitter, i.e. signal launched into one of the ports is distributed symmetrically and equally between the two output ports of the junction. We search for the optimum regime of beam splitter operation in order to maximize the output power and reduce the radiation losses. For this reason we scan over the wavelength range near the structure resonance, and by varying the junction parameters we trace for the maximum power transmission. Following this procedure we find that the optimum width of waveguides is $a=350$nm and their permittivity is $\varepsilon=3$. For matching purposes, we also introduce a $0.9 \mu$m-long transition waveguides with permittivity $3.5$ (see Fig.~\ref{geometry}) in order to reduce backscattering and increase the transmission. We note that by placing the structure into the core of the junction its resonance is shifted to  $1.43 \mu$m due to a change in boundary conditions.

In Fig.~\ref{circulat}(a) we plot the simulation results for the z-oriented magnetic field distribution at the resonance for a signal launched into port 1. Clearly a symmetric beam splitting into the other two output arms is observed. In order to trace the operation efficiency we study the normalized power transmission $S_{21(31)}=P_{out}^{2(3)}/P_{in}^{1}$ with frequency variation, see Fig.~\ref{charact}(a). We observe a strong transmission resonance at free-space wavelength $\lambda\simeq1.43 \mu$m, which corresponds to the plasmonic resonance of the entire structure. At the resonance, the power output at each of the ports 2 and 3 ($S_{21}=S_{31}$) reaches up-to $33\%$ and then steeply decreases with frequency detuning. The strong dependence and correlation between the power output and the structure excitation reveals a crucial role of structure resonance in the junction operation. The structure being an efficient plasmonic scatterer when excited redistributes the input power between the junction arms, i.e. ``directionally" scatters the input signal and thus increasing the out-coupling efficiency. Off resonance the scatterer is not excited as much, and therefore the system behaves as a conventional photonic Y-junction splitter, which is rather inefficient due to a large angle between the junction arms so that most of the input power is radiated away. We note that, despite of significant increase in transmission with the use of the nanostructure at its resonance, the radiation losses remain quite high; however we believe that further geometry optimization, including the redesign of the plasmonic structure itself, will increase the performance.

\begin{figure}[h!]
  \begin{center}
      \includegraphics[width=1\columnwidth]{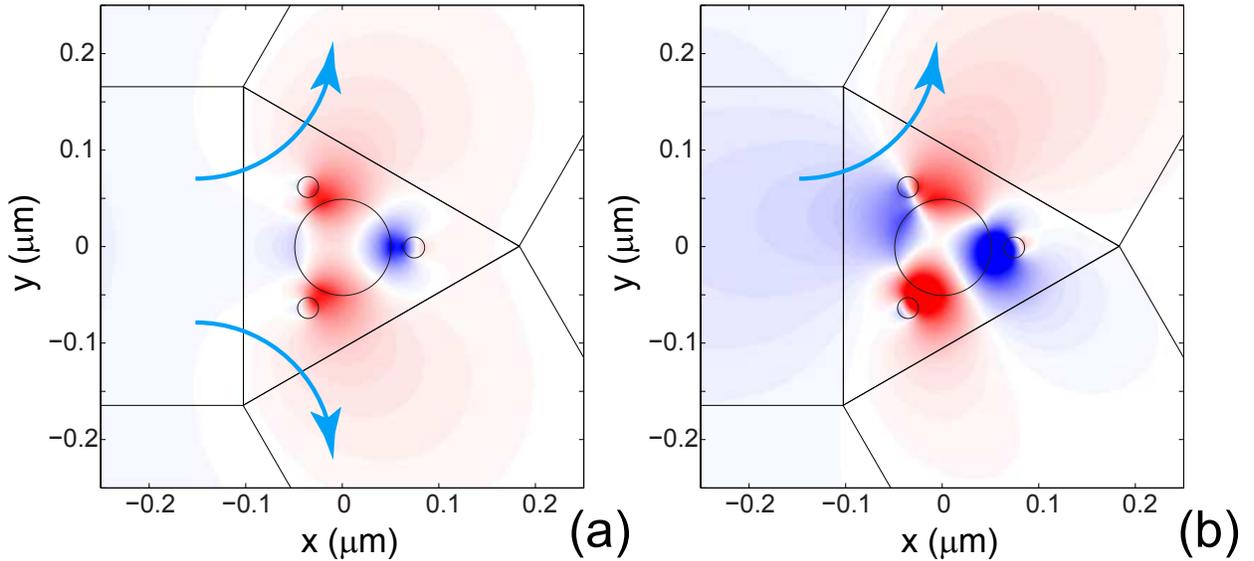}
      \caption{Snapshot in time of the near-field distribution of the magnetic field ($H_z$) excited at the resonance of the plasmonic structure ($\lambda = 1.43\mu$m) for a) unbiased ($\alpha=0$) and b) biased ($\alpha=0.1$) systems, respectively. The arrows show schematically the coupling between input port and the two output ports.} \label{nearfield}
  \end{center}
\end{figure}

As the next step, we introduce the MO activity ($\alpha\neq0$) and study its influence on the circulator operation. Fig.~\ref{circulat}(b) shows the numerical results for the z-oriented magnetic field distribution for the signal launched into the port 1 at the resonance frequency ($\lambda = 1.43 \mu$m) when $\alpha=0.1$. In this case we observe a substantial breaking of the symmetry between the output at ports 2 and 3. The power is predominantly guided into the arm 2, while the other arm carries much less power. To reveal the mechanism of such strong symmetry breaking we study the near-field distributions for biased and unbiased systems, see Fig.~\ref{nearfield}(a,b). The near-field pattern in the unbiased case, Fig.~\ref{nearfield}(a), is symmetric and resembles that of a ``point" magnetic quadrupole. Such near-field distribution ensures a strong and symmetric coupling between the junction arms and the structure. When MO activity is introduced the nearfield radiation is tilted and rotated~\cite{our_prl}. This rotation of the scattering pattern  leads to the change of coupling between the structure and the junction arms, so that input (port 1) and only one of the output (port 2) waveguides are coupled efficiently with the structure. Furthermore, we notice from Fig.~\ref{nearfield}(b) that a predominantly antisymmetric field profile is induced at the entrance of one of the output arms (port 3 in this case), which is why the single-mode waveguide is not excited, causing strong isolation of the corresponding port, as shown on Fig.~\ref{circulat}(b). We note that the suggested mechanism of circulator operation based on the scattering and the rotation of the scattering differs conceptually from the well established interference principle employed in the conventional circulator designs~\cite{Helszajn_1}.

\begin{figure}[h!]
  \begin{center}
      \includegraphics[width=1\columnwidth]{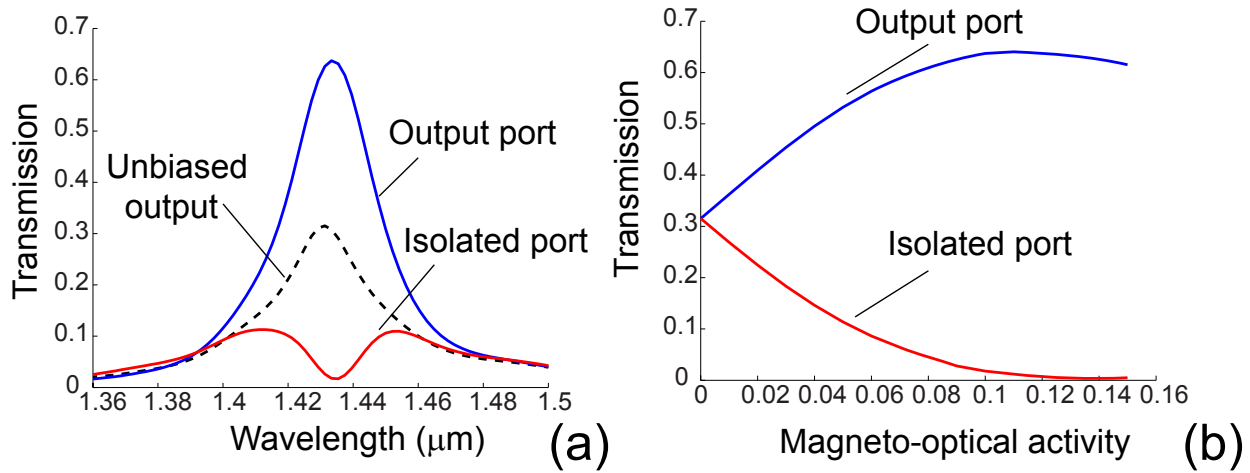}
      \caption{a) Variation of the circulator power transmission into the port 2 and isolation of the port 3, as a function of wavelength for biased ($\alpha=0.1$) and unbiased ($\alpha=0$) cases. b) Variation of the circulator power transmission as a function of MO activity parameter at the free space wavelength ($\lambda=1.43\mu$m). Here metal permittivity is assumed to be frequency indedpendent and fixed at $\varepsilon_m = -10$.} \label{charact}
  \end{center}
\end{figure}

Next we trace the circulator performance as a function of wavelength when $\alpha=0.1$, see Fig.~\ref{charact}(a). We observe again a resonant behavior which clearly manifests the role of the structure in the circulator operation. At resonance we find that the power transmitted to port 2 comprises up-to $63\%$ of the input power, whereas less than $3\%$ is transmitted from port 1 to port 3 demonstrating  a good isolation of the corresponding port. We note that $63\%$ of power output observed in the biased case is in fact about $96\%$ of total power output in the unbiased case ($33\%+33\%$), meaning that the radiation losses in the biased regime stay almost the same as in the unbiased one.

In Fig.~\ref{charact}(b) we plot the transmission at ports 2 and 3 as a function of MO activity parameter $\alpha$ at $\lambda =  1.43 \mu$m. We observe a growth of power output for the port 2 with a corresponding increase of isolation for the port 3. The transmission through port 2 peaks with maximum of $64\%$ when $\alpha = 0.11$ and begins to decline with a subsequent increase of $\alpha$. For small values of $\alpha$ we observe linear dependence of transmission and isolation on $\alpha$ which correlates well with the empirical formula describing the frequency splitting between the counter rotating states. The deviation from the linear dependence with growth of $\alpha$ can be explained by slight shift of resonance frequency and the change of coupling efficiency between the output waveguides and the structure due an ``over rotation'' of the near-field pattern. We note that at $\alpha=0.06$
corresponding to the realistic levels achieved in BIG we observe about $54\%$ of transmission from port 1 to port 2 and less than $10\%$ from port 1 to port 3.

In order to underline the nonreciprocal nature of the proposed plasmonic circulator we launch signal from port 2. In this case we clearly see that light is emitted from port 3, whereas port 1 is isolated, as expected. Hence, the device when biased transmits the signal from any port to the next port (in circular order). Changing the direction of magnetization will swap the output and isolated ports, as also expected. Note also that the design principle we utilized based on the scattering from the structure provides the potential for the design of a circulator with higher number of ports.

\begin{figure}[h!]
  \begin{center}
      \includegraphics[width=0.8\columnwidth]{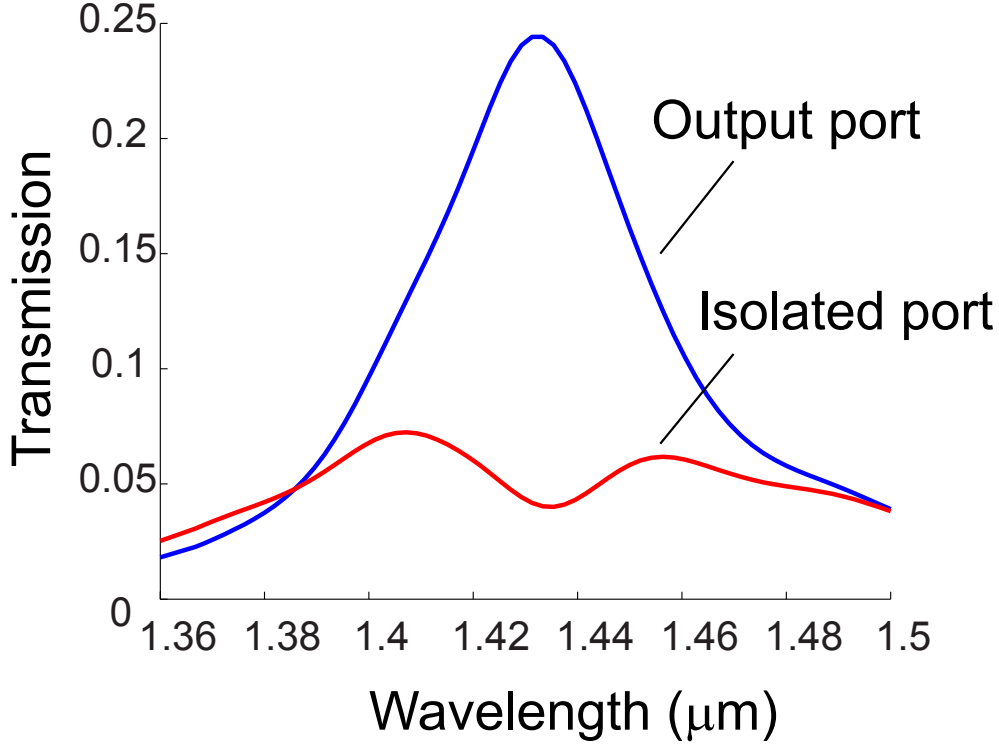}
      \caption{The effect of metal losses on the circulator performance. Power transmission as a function of wavelength for $\alpha=0.1$ is plotted for the output and isolated ports. Here metal permittivity is assumed to be frequency indedpendent and fixed at $\varepsilon_m = -10-0.01i$.} \label{lossy}
  \end{center}
\end{figure}

Finally, we study the influence of losses on the performance of the proposed circulator. For this purpose we consider the metal to be lossy, described by a complex permittivity $\varepsilon_m = -10-0.01i$. In Fig.~\ref{lossy} we plot the transmission as a function of frequency for both output and isolated ports when $\alpha=0.1$. Clearly the symmetry breaking in transmitted power between the output ports is still observed. Again we achieve resonant transmission though one of the ports only, while the other output port is isolated. However, as expected the losses result in decrease in the total power transmission, and a decrease in the isolation ratio, i.e. the ratio of transmitted powers through both output ports. The ability to keep the circulator performance with losses allows the possibility of implementing ferromagnetic metals possessing stronger MO activity (such as, for example, Co, Ni, Fe). Although these materials exhibit quite high electromagnetic damping, their combination with plasmonic nobel metals and proper structure engineering might give the possibility to control the effects of the losses and the MO activity~\cite{Wang_coreshell,Jain_coreshell}.

In conclusion, we have proposed a design for a subwavelength three-port junction circulator with an integrated plasmonic structure. We have demonstrated numerically that the plasmonic resonance and the rotation of corresponding near-field scattering in presence of MO activity leads to the breaking of symmetry in the power output between the circulator arms. Our analysis reveals up-to  $63\%$ of power transmitted through one of the output ports, with almost complete isolation for the other output port of the circulator. Finally, we have studied the influence of losses and found that despite the decrease of net output in lossy systems, the symmetry breaking between the output ports remains relatively high. This concept may find useful application in integrated nanophotonics, optical metatronics, and signal handling in on-chip integrated circuits.

\section{Acknowledgements}
Authors thank A. Mahmoud for useful discussion. This work is supported in part by the US Air Force Office of Scientific Research (AFOSR) grant number FA9550-10-1-0408.

\bibliography{achemso}

\end{sloppy}
\end{document}